\documentstyle[preprint,aps]{revtex}
\begin{document}
\draft
\preprint{CLNS 97/1525}
\title{Aspects of Quenched Chiral Perturbation Theory\\for Matter Fields}
\author{Chi--Keung Chow}
\address{Newman Laboratory of Nuclear Studies, Cornell University, Ithaca,
NY 14853.}
\date{\today}
\maketitle
\begin{abstract} 
Three topics about the application of quenched chiral perturbation theory 
to matter fields are studied.  
It is proved that the hairpin axial current couplings in quenched chiral 
perturbation theories do not contribute to the quenched chiral singularities 
for one chiral loop renormalization of matter field properties.  
The modification of mass corrections in the chiral limit due to nonzero mass 
splittings are studied, and selection rules for hadron decays in quenched 
QCD are obtained.  
\end{abstract}
\narrowtext
Most lattice QCD simulations are performed under the quenched approximation, 
where the fermionic determinant is set to unity.  
While in the real world the $u$ and $d$ quarks have masses much less than 
$\Lambda_{\rm QCD}$, lattice QCD are often simulated with $m_q \sim 
\Lambda_{\rm QCD}$, and physical results are extracted by extraplation to the 
chiral limit $m_q\to 0$.  
As a result, it is important to understand the chiral limit of quenched 
QCD.  
Over the past few years, quenched chiral perturbation theory (Q$\chi$PT) 
has been formulated as the low energy effective theory of quenched QCD 
\cite{Morel,S1,S2}.  
In Q$\chi$PT, the cancelation of internal quark loops is enforced by the 
introduction of ghost quarks (quarks with wrong -- bosonic -- statistics) 
\cite{BG1}.  
Since the contributions of ghost loops to path integrals are equal in 
magnitude but opposite in sign to those of quark loops, there will be a 
complete cancelation of loop contribution if there are as many quarks as 
ghosts, with the same masses.  
(One may also consider partially quenched chiral perturbation theories 
\cite{BG2}, in which the number of quarks and the number of ghosts are 
different.)  
Just as in ordinary chiral perturbation theory ($\chi$PT) the low energy 
dynamics is dominated by the Goldstone bosons (pions for two flavor QCD) 
which are quark-antiquark bound states, in Q$\chi$PT the low energy dynamics 
is dominated by the Goldstone ``bosons'' which may be quark-antiquark, 
ghost-antighost, quark-antighost or ghost-antiquark bound states.  

The main difference between $\chi$PT and Q$\chi$PT lies in the the role 
played by the $\eta'$ meson.  
In ordinary QCD, $\eta'$ acquires a heavy mass due to the necklace diagrams 
and hence are integrated out in the standard formulation of $\chi$PT.  
Since the necklace diagrams involve quark loops, however, they do not 
contribute in the quenched theory.  
As a result, the $\eta'$ have the same mass as the other Goldstone bosons 
(massless in the chiral limit) and cannot be integrated out.  
It does differ from the other Goldstone bosons in having an extra term in 
its propagator due the the hairpin diagram.  
\begin{equation}
{1\over p^2-m^2}\pmatrix{1&0\cr 0&-1\cr}+{-nA_0 p^2 + nM_0^2\over(p^2-m^2)^2}
\pmatrix{1&1\cr1&1\cr}, 
\end{equation}
where the first row/column represents the $\eta'$ meson and the second 
row/column represents the $\tilde \eta'$, the ghost-antighost analog of 
$\eta'$, and $n$ is both the number of quark flavors and the number of ghost 
flavors.  
The first term is the usual propagator where $\eta'$ and $\tilde \eta'$ 
propagate independently with a single pole at $p^2 = m^2$, which denotes both 
$m_{\eta'}$ and $m_\pi$ as $\pi$ and $\eta'$ are degenerate in quenched QCD.  
The second term carries a double pole and represents a possible mixing 
between $\eta'$ and $\tilde \eta'$, characterized by the parameters $A_0$ and 
$M_0^2$.
\footnote{Here we are considering quenched QCD with two light flavors and we 
have set $m_u = m_d$.  
With three light flavors and $m_s \neq m_q$, the form of the double pole term 
will be more complicated.  
However, our analysis is independent of the particular form of the double 
pole term.}

Besides the Goldstone boson fields, one can also study the dynamics of matter 
fields (vector and tensor mesons \cite{BCF,CR1}, baryons \cite{LS,KK,CR2}, 
heavy mesons \cite{B,SZ,SZ2}, heavy baryons \cite{C}, {\it etc.}) under the 
framework of Q$\chi$PT.  
In this paper, we will discuss three theoretical issue around the application 
of Q$\chi$PT on matter fields.  
The first section will prove the (ir)relevance of hairpin axial couplings 
(which will be defined below) to quenched chiral corrections of matter 
field quantities.  
Then the effects of mass splittings in Q$\chi$PT will be discussed, followed 
by the last section which study the possibility of understanding hadron decay 
in Q$\chi$PT.  

\section{Hairpin Axial Couplings and Quenched Chiral Corrections}

Since the matter fields are significantly heavier than $\Lambda_{\rm QCD}$, 
it is customary to introduce the matter fields as heavy particles, {\it i.e.},
static fields with definite velocities.  
Under the heavy particle formalism, the propagators of the matter fields are 
always $(v\cdot p)^{-1}$, and in the leading order of a derivative expansion 
the matter fields couples to the Goldstone bosons through the axial currents.  
\begin{equation}
{\cal L}_{\rm standard} = i g \,{\rm Str}\,[X^\mu_{(H_{(f)}^\dag, H_{(i)})} 
A_\mu], 
\end{equation}
where $A_\mu = {i\over2}(\xi\partial_\mu\xi^\dag - \xi^\dag\partial_\mu\xi)$ 
is the Goldstone boson axial current, ``Str'' denotes the supertrace over 
all the quark and ghost flavors, and $X^\mu$ is the axial current operator 
between the initial state $H_i$ and the final state $H_f$.  
The form of $X^\mu$ depends on the matter fields in question.  
For example, the axial current of the vector meson fields $V^\lambda$ takes 
the following form \cite{BCF,CR1}: 
\begin{equation}
X^\mu_{({V^\lambda_{(f)}}^\dag, V^\sigma_{(i)})} = 
\epsilon^{\mu\nu\lambda\sigma} v_\nu 
\{{V_\lambda}_{(f)}^\dag, {V_\sigma}_{(i)}\}, 
\end{equation}
where $g$ is an undetermined coupling constant, while the exact forms of the 
baryon axial current couplings can be found in Ref.~\cite{LS}.  

The chiral lagrangian of Q$\chi$PT is complicated by the existence of 
hairpin axial current couplings, which couplings $\eta'$ and $\tilde \eta'$ 
to the flavor singlet axial current of matter fields.  
\begin{eqnarray}
{\cal L}_{\rm hairpin} &=& i h \,{\rm Str}\,[X^\mu_{(H_{(f)}^\dag, H_{(i)})}]
{\rm Str}[A_\mu]\nonumber\\ &=&i n^{1/2} (h/f) \,{\rm Str}\,
[X^\mu_{(H_{(f)}^\dag, H_{(i)})}]\partial_\mu (\eta' - \tilde \eta').  
\end{eqnarray}

After writing down the interaction lagrangians, one can study the chiral 
limit of different physical quantities by calculating the one chiral loop 
renormalizations.  
For example, the mass renormalization of the matter fields can be evaluated 
through the following integral, 
\begin{eqnarray}
\Delta m &=& \int {d^4p\over(2\pi)^4} \left[{(\hbox{coup.~const.})^2\over
f^2} \sum_f X^\nu_{(H_{(i)}^\dag, H_{(f)})} X^\mu_{(H_{(f)}^\dag, H_{(i)})}
\right] \nonumber\\ &&\qquad{p_\nu p_\mu\over v\cdot p}\;
(\hbox{propagator of the Goldstone boson}).  
\end{eqnarray}
where the sum is over all possible intermediate states.  
Note that the expression inside the square brackets are independent of $p$ 
and hence can be taken out of the integral.  
So $\Delta M$ just depends on the form of the propagators of the Goldstone 
bosons.  
For all the flavor non-singlet states the propagators are $(p^2-m^2)^{-1}$, 
which is also the form of the ``single pole'' propagator of $\eta'$ and 
$\tilde \eta'$.  
Putting these propagators in the integral with contribute to $\Delta M$ 
with the generic form 
\begin{equation}
\Delta M \sim \int {d^4p\over(2\pi)^4}{p^2\over v\cdot p}{1\over p^2-m^2}
\sim m^3.  
\end{equation}
This $m^3 \sim m_q^{3/2}$ term has a singular dependence of $m_q$ which is 
due to low-energy chiral dynamics and cannot be canceled by counterterms.  
Since such a term also appear in normal $\chi$PT, they are called standard 
chiral singularities.  
In contrast, if the contribution to the loop integral due to the ``double 
pole'' propagator is of the form:
\begin{equation}
\Delta M \sim \int {d^4p\over(2\pi)^4}{p^2\over v\cdot p}
{-A_0p^2 + M_0^2\over (p^2-m^2)^2}\sim M_0^2 m.  
\end{equation}
This term $M_0^2 m\sim m_q^{-1/2}$ is more singular in the chiral limit than 
the $m^3$ term above and is a quenched artifact, which is usually called 
quenched chiral singularities.  
Since these quenched chiral singularities dominate over the standard ones 
in the chiral limit, understanding them may be helpful in the determination 
of the best extrapolation of lattice QCD results to the zero mass limit.  

Naively, both the standard and quenched chiral singularities may depend on 
the normal axial coupling $g$ and the hairpin axial coupling $h$.  
It turns out that the quenched chiral singularities are independent of the 
value of $h$.  
The reasoning is very simple: the quenched chiral singularities come from 
$\eta'$ and $\tilde \eta'$ loops, and these Goldstone bosons can couple to 
the matter field through either the standard or hairpin axial current 
couplings.  
\begin{mathletters}
The matter field, made entirely of quarks and/or antiquarks, cannot couple to 
$\tilde \eta'$ through the standard axial coupling, hence
\begin{equation}
g_{{\rm standard}\,\eta'} = n^{1/2} g,\qquad 
g_{{\rm standard}\,\tilde \eta'} = 0.  
\end{equation}
On the other hand, $\eta'$ and $\tilde \eta'$ couple at the same strength 
through the hairpin coupling, but with opposite signs.  
\begin{equation}
g_{{\rm hairpin}\,\eta'} = n^{1/2} h,\qquad 
g_{{\rm hairpin}\,\tilde \eta'} = -n^{1/2} h.  
\end{equation}
Hence the total couplings are 
\begin{equation}
g_{{\rm total}\,\eta'} = n^{1/2} (g+h),\qquad 
g_{{\rm total}\,\tilde \eta'} = -n^{1/2} h.  
\end{equation}
\end{mathletters}
The one loop integral is proportional to 
\begin{eqnarray}
n(g_{{\rm total}\,\eta'}g_{{\rm total}\,\eta'}&&+g_{{\rm total}\,\eta'}
g_{{\rm total}\,\tilde \eta'}+g_{{\rm total}\,\tilde \eta'}
g_{{\rm total}\,\eta'}+g_{{\rm total}\,\tilde \eta'}
g_{{\rm total}\,\tilde \eta'})\nonumber\\&&=n((g+h)^2+2(g+h)(-h)+(-h)^2)
=g^2n.  
\end{eqnarray}
All $h$ dependences have canceled, as claimed.  

Actually, the result can be obtained in a more elegant way.  
Let's define $\eta'_\pm = \eta' \pm \tilde \eta'$, where $\eta'_+$ and 
$\eta'_-$ are orthogonal linear combinations.  
Notice that the hairpin axial couplings always couple to Str$(A_\mu) = 
\partial_\mu(\eta' - \tilde \eta') = \partial \eta'_-$.  
On the other hand, since the ``double pole'' propagator mixes $\eta'$ and 
$\tilde \eta'$, one can diagonalize it to find out its eigenstates.  
They turn out to be exactly $\eta'_\pm$.  
\begin{equation}
\bordermatrix{&\eta'&\tilde \eta'\cr \eta'&1&1\cr \tilde \eta'&1&1\cr} 
\longrightarrow\bordermatrix{&\eta'_+&\eta'_-\cr\eta'_+&2&0\cr\eta'_-&0&0\cr} 
\end{equation}
Since $\eta'_-$ has eigenvalue zero, it does not propagate under the ``double 
pole'' propagator at all!  
In other words, quenched chiral singularities are solely due to $\eta'_+$ 
loops, while the hairpin axial couplings couple only to $\partial \eta'_-$.  
As a result, the hairpin axial current couplings cannot contribute to 
the quenched chiral singularities.  

After proving that the hairpin axial current couplings do not contribute to 
quenched chiral singularities for mass renormalizations, it is time to 
study the possible generalizations and/or limitation of our result.  
First of all, can our analysis be extended to more than two flavors, with 
possibly different masses for different flavors?  
In the presence of different quark masses, it is more convenient to use 
instead of the singlet basis $(\eta', \tilde \eta')$, the flavor basis 
$(u\bar u, \tilde u\skew2\bar{\tilde u}, d\bar d, \tilde d\skew6\bar{\tilde d},
s\bar s, \tilde s\skew2\bar{\tilde s}, \dots)$, where $\tilde q$ is the ghost 
degenerate with quark $q$.  
The hairpin axial current couplings go through $\partial_\mu \eta_- = 
\partial_\mu \sum_{k=1}^n (q\bar q - \tilde q \skew3\bar {\tilde q})$, in the 
leading order of flavor symmetry breaking.  
On the other hand, the ``double pole'' propagator now takes the form 
$(-A_0 p^2 + M_0^2) {\cal M}$, where ${\cal M}$ is a $2n \times 2n$ matrix.  
\begin{eqnarray} 
{\cal M}&=&\bordermatrix{&u\bar u& \tilde u \skew2\bar{\tilde u}& d\bar d& 
\tilde d \skew6\bar{\tilde d}&s\bar s& \tilde s \skew2\bar{\tilde s}\cr u\bar u
&P_uP_u&P_uP_u&P_uP_d&P_uP_d&P_uP_s&P_uP_s\cr \tilde u \skew2\bar{\tilde u}
&P_uP_u&P_uP_u&P_uP_d&P_uP_d&P_uP_s&P_uP_s\cr d\bar d
&P_dP_u&P_dP_u&P_dP_d&P_dP_d&P_dP_s&P_dP_s\cr \tilde d \skew6\bar{\tilde d}
&P_dP_u&P_dP_u&P_dP_d&P_dP_d&P_dP_s&P_dP_s\cr s\bar s
&P_sP_u&P_sP_u&P_sP_d&P_sP_d&P_sP_s&P_sP_s\cr \tilde s \skew2\bar{\tilde s}
&P_sP_u&P_sP_u&P_sP_d&P_sP_d&P_sP_s&P_sP_s\cr} \nonumber\\ &=&
\bordermatrix{&u\bar u&d\bar d&s\bar s\cr u\bar u&P_uP_u&P_uP_d&P_uP_s\cr
d\bar d&P_dP_u&P_dP_d&P_dP_s\cr s\bar s&P_sP_u&P_sP_d&P_sP_s\cr} \otimes 
\bordermatrix{&q\bar q&\tilde q\skew3\bar{\tilde q}\cr q\bar q&1&1\cr 
\tilde q\skew3\bar{\tilde q}&1&1\cr}, 
\end{eqnarray}
where $P_q = (p^2 - m_{q\bar q}^2)^{-1}$.  
It is now easy to check that $\cal M$ annihilate $\eta'_-$, especially if 
one notice that by the same representation $\eta'_-$ can be expressed as 
\begin{equation}
\eta'_- = \pmatrix{1\cr-1\cr1\cr-1\cr1\cr-1\cr} = \pmatrix{1\cr1\cr1\cr} 
\otimes \pmatrix {1\cr-1}, 
\end{equation}
and the last two dimensional column submatrix is annihilated by the 
$2\times 2$ square submatrix in $\cal M$.  
Hence our result holds even when the quarks are not degenerate.  
In fact it only depends on the one-to-one degeneracy between the quarks and 
the ghosts.  
It is also clear in this representation that our result {\it cannot\/} be 
generalized to partially quenched theories, {\it i.e.}, theories in which 
the number of quarks and ghosts are different.  
In fact it is straightforward to show that, in partially quenched theories 
with $n$ quarks and $k$ ghosts, the quenched chiral singularities \footnote{
In partially quenched chiral perturbation theories, the $\eta'$ and $\tilde 
\eta'$ mesons do not have double pole in their propagators.  
Instead they have two different single poles, one at the original position 
$p^2=m^2$, while the other is shifted to $p^2=m^2+\delta m^2$.  
Please refer to Ref.~\cite{BG2} for details.  
Here we refer to the renormalization due to the shifted pole propagator as 
quenched chiral singularities, but in fact they are no more singular than 
the standard corrections unless one goes to the fully quenched limit.} 
have the generic form $(ng + (n-k)h)^2$, where $g$ and $h$ are the standard 
and hairpin axial current couplings respectively.  

So far we have been studying exclusively mass renormalization.  
Does our result hold for renormalization of other physical quantities?  
Before answering this question, we must ascertain what we mean by standard 
and/or quenched chiral singularities under such circumstances.  
Suppose we are interested in studying the chiral correction to a certain 
operator $\cal O$ which acts on matter fields.  
The ``single pole'' propagator will lead to operator renormalization with the 
following integral form: 
\begin{equation}
Z_{\cal O} = \int {d^4p\over(2\pi)^4} \left[{(\hbox{coup.~const.})^2\over
f^2} \sum_f X^\nu_{(H_{(i)}^\dag, H_{(f)})} X^\mu_{(H_{(f)}^\dag, H_{(i)})}
\right]{p_\nu\over v\cdot p}{\cal O}{p_\mu\over v\cdot p} {1\over p^2-m^2}.  
\end{equation}
We will refer whatever chiral correction induced by this term as standard 
chiral corrections.  
On the other hand, the ``double pole'' will induce corrections of another 
form: 
\begin{equation}
Z_{\cal O} = \int {d^4p\over(2\pi)^4} \left[{(\hbox{coup.~const.})^2\over
f^2} \sum_f X^\nu_{(H_{(i)}^\dag, H_{(f)})} X^\mu_{(H_{(f)}^\dag, H_{(i)})}
\right]{p_\nu\over v\cdot p}{\cal O}{p_\mu\over v\cdot p} 
{-nA_0 + nM_0^2 \over (p^2-m^2)^2}, 
\end{equation}
which we will refer as quenched chiral corrections.  
Generically, if the standard chiral correction takes the form of $m^\alpha\sim
m_q^{\alpha/2}$ in the chiral limit, the quenched chiral correction is more 
singular with the form $M_0^2 m^{\alpha-2}\sim m_q^{\alpha/2-1}$.  
It is straightforward to generalize our result to the following statement: 
{\sl Quenched one-chiral loop corrections for matter field properties are 
independent of the values of hairpin couplings.}
This is the main result of this section, and there is one important exception 
to this rule, which will be discussed later.  

The author does {\it not\/} claim to be the first to note this pattern of 
cancelation of hairpin axial coupling in quenched chiral divergences.  
Our result is ``immediate'' (to quote Steve Sharpe \cite{PC}) from the quark 
flow diagrams in Ref.~\cite{SZ2}.  
When one surveys the existing literature it is probable that the authors of 
previous articles on Q$\chi$PT on matter field are aware of such a pattern 
(although, as far as the author is aware of, this is the first systematic 
discussion of this result).  
For example, Labrenz and Sharpe has calculated the quenched chiral 
singularities (Eq.~(97 -- 99) in Ref.~\cite{LS}) for baryons in three flavor 
Q$\chi$PT and indeed they are independent of the the hairpin axial couplings. 
So are the quenched chiral corrections to the octet baryon axial charges as 
calculated by Kim and Kim \cite{KK}.  
Our rule also holds for heavy mesons, where Booth \cite{B} as well as Sharpe 
and Zhang \cite{SZ,SZ2} has calculated the renormalizations of heavy meson 
decay constants, masses, bag constants, and Isgur--Wise form factors.  
In all these cases the hairpin axial couplings do not contribute to the 
quenched chiral corrections.  
The same can be said for Chiladze's work on heavy baryons \cite{C}.  
For easy comparison with the literature we list the standard and hairpin 
axial coupling constants in Ref.~\cite{BCF,CR1,LS,KK,CR2,B,SZ,SZ2,C} in the 
following table: 

\bigskip
\bigskip

\centerline{\vbox{
\halign{\hfil#\hfil\qquad&\hfil#\hfil&\qquad\hfil#\hfil&\qquad\hfil#\hfil\cr
Matter Fields&Reference&Standard&Hairpin\cr
\noalign{\smallskip}
vector mesons&Booth, Chiladze, Falk \cite{BCF}&$g_2$&$g_4$\cr
tensor mesons&Chow, Rey \cite{CR1}&$\tilde g_2$&$\tilde g_4$\cr
baryons $(N_c=3)$&Labrenz, Sharpe \cite{LS}&$D$,$F$,$\cal H$,$\cal C$&
$\gamma$,$\gamma'$\cr
baryons $(N_c=3)$&Kim, Kim \cite{KK}&$D$,$F$,$H$,$C$&$\gamma$,$\gamma'$\cr
baryons $(N_c\to\infty)$&Chow, Rey \cite{CR2}&$g$&not considered\cr
heavy mesons&Booth \cite{B}&$g$&$\gamma$\cr
heavy mesons&Sharpe and Zhang \cite{SZ,SZ2}&$g$&$g'$\cr
heavy baryons&Chiladze \cite{C}&$g_2$,$g_3$&$g_1$\cr
}}}

\bigskip

More interesting are the cases where this result actually fails.  
This is the case for the paper by Booth, Chiladze and Falk on Q$\chi$PT for 
vector mesons \cite{BCF}, which is subsequently extended to the tensor mesons 
case by Chow and Rey \cite{CR1}.  
For these cases the hairpin axial current couplings, which is denoted as 
$g_4$ in Ref.~\cite{BCF}, does contribute to the quenched chiral 
singularities.  
The reason for this breakdown of our result is clear.  
In our analysis above we have assumed that the only source of quenched 
chiral singularities is the ``double pole'' term in the $\eta'$ and $\tilde 
\eta'$ propagators, while for the case of vector and tensor mesons there 
are additional sources of quenched singularities, namely the possible 
hairpin diagrams for these matter fields themselves, which may lead to 
``double pole'' propagators for these matter fields.  
However, if one explicitly setting the hairpin diagrams on these matter 
fields to be zero ({\it i.e.}, if one set the parameters $\mu_0=0$ and 
$A_N=0$ in Ref.~\cite{BCF}), then our results are applicable.  
In other words, the hairpin axial current coupling $g_4$ cannot contribute 
to quenched chiral singularities which involve the $\eta'$ and $\tilde \eta'$ 
``double pole'' propagators ($I_2$ and $I_4$ in Ref.~\cite{BCF}) but in 
general will contribute to the standard chiral singularities ($I_1$) as well 
as the quenched chiral singularities which do not involve the $\eta'$ ``double 
pole'' propagator ($I_3$).  
Also note that there are more hairpin couplings for Q$\chi$PT for vector and 
tensor mesons, which are called $g_1$ and $g_3$ in Ref.~\cite{BCF}.  
The definition of $g_3$ does involve Str$[A_\mu] \sim \partial_\mu \eta'_-$ 
and hence our result is applicable to this $g_3$ coupling.  
Indeed it does not contribute to any of the quenched chiral singularities.  
On the other hand, the definition of $g_1$ does {\it not\/} involve 
Str$[A_\mu]$ and hence is not constrained by our result.  

In summary, we have proved that the quenched chiral corrections to matter 
field properties are independent of the values of the hairpin axial current 
couplings.  
Besides serving as a useful check of the correctness of Q$\chi$PT 
calculation results, this observation may be useful in the eventual 
determination of parameters in the Q$\chi$PT lagrangians.  
All the coupling constants appearing in the Q$\chi$PT lagrangians are 
nonperturbative quantities which cannot be calculated from first principles 
from quenched QCD and has to be extracted from lattice results (just as 
the paramaters in standard $\chi$PT lagrangians have to be extracted from 
experimental data).  
While one can extract the ``double pole'' parameters from lattice results 
on Goldstone boson systems, it may be difficult to disentangle the effects 
of the standard axial couplings from their hairpin counterparts.  
Our result suggests the procedure that one should extract the values of the 
standard axial couplings from the quenched chiral corrections, and with 
those pieces of information one can proceed to extract the values of the 
hairpin axial couplings from the standard chiral divergences.  

\section{Mass Splitting and Chiral Mass Corrections}

Chiral perturbation theory, both the standard and quenched versions, provides 
a systematic framework to describe low energy interactions between hadrons.  
Before one constructs a chiral lagrangian, however, one must decide which 
fields to include.  
There is no clear cut criteria on which fields should be included, but the 
conventional wisdom suggested that one should {\it at least\/} include all 
fields which are related by light flavor SU($n$) ($n=2$ or 3, depending on the
scenario), as well as all the fields which are degenerate in the large $N_c$ 
limit.  
If the hadron in question contains a heavy quark, then we should also include 
all the fields related by heavy quark spin symmetry as well.  
For example, in Ref.~\cite{BCF} both the vector meson octet and the singlet 
appeared in the chiral lagrangian, as the octet and the singlet are 
degenerate in the large $N_c$ limit.  
In fact they can be conveniently combined into a U(3) nonet.  
Similarly in Ref.~\cite{LS} the baryon chiral lagrangian contains both the 
octet and the decuplet, which are again degenerate in the large $N_c$ limit.  
It is well known that, in standard $\chi$PT, the failure to include the 
baryon decuplet will lead to huge chiral corrections to baryon octet, while 
the chiral loops in a $\chi$PT with both the octet and the decuplet is much 
smaller \cite{JM}.  
Physically these degenerate states are strongly coupled to each other, and 
the failure to include them will lead to an incomplete representation of the 
interaction, causing a badly-behaved loop expansion.  

In both the real world and its quenched approximation on the lattice, however, 
all these symmetries (chiral, heavy quark, and large $N_c$) are broken and 
the degeneracies are only approximate.  
In (Q)$\chi$PT, one often (though not always) treats these mass splittings as 
small perturbations and include them leading to the leading order.  
While such approximations are usually justified, they are not entirely 
satisfactory as it is not clear if the higher order effects will change the 
chiral behavior drastically.  
The question can be put more impendingly in the following way.  
Let's take the chiral corrections to hadron masses as an example.  
Since the one loop diagram has two axial current couplings, each giving a 
factor of $1/f$, one can obtain from dimensional analysis that 
\begin{equation}
\Delta M \sim m^3/f^2, 
\end{equation}
simply on the basis that the Goldstone boson mass $m$ is {\it the only other 
mass scale in the theory}.  \footnote{Note that chiral logarithms like $m^Z 
\ln {m^2\over \mu^2}$ appear only when $Z$ is an integer, so that the effect 
of shifting the subtraction point $\mu$ can be canceled by a counterterm 
analytic in $m_q\sim m^2$.  In this case $Z=3$, and chiral logarithms are 
forbidden.}  
In other words, the chiral behavior of the mass correction cannot be changed 
unless some other mass scale is introduced in the theory.  
In Q$\chi$PT, the new mass scale is $M_0^2$, and the quenched chiral 
singularities are proportional to $M_0^2m$, which dominate in the chiral 
limit.  
Now the mass splitting $\Delta$ between states is another new mass scale and 
can potentially change the chiral behavior.  
The question is, will it?  

Fortunately, it is possible to study the Feynman integral exactly.  
This has been done in $\chi$PT in Ref.~\cite{JM} for baryons (though the 
treatment is general and can be applied to other systems as well), and the 
generalization to Q$\chi$PT is trivial.  

Let's review the analysis in Ref.~\cite{JM}, which study the integral 
\footnote{There is a factor of $i$ difference between our definition of $I$ 
and that in Ref.~\cite{JM}.}
\begin{equation}
I = {i\over f^2}\int {d^4p\over(2\pi)^4} {p^2\over v\cdot p - \Delta}
{1\over p^2 - m^2}, 
\end{equation}
where we will assume $\Delta > 0$ so that the intermediate state is heavier 
than the external state and hence the integral does not have any cuts 
corresponding to real particle decays (which will be the subject matter of the 
following section).  
Then the integral can be evaluated exactly.  
It is instructive to separate the result into two parts: 
\begin{equation}
I = I^{\rm (a)} + I^{\rm (na)}
\end{equation}
The analytic contribution 
\begin{equation}
I^{\rm (a)} = {1\over 12\pi f^2} ({5\over3\pi}\Delta^3 + {2\over\pi}\Delta m^2)
\end{equation}
can be canceled by counterterms, while the nonanalytic contribution cannot be 
modified by counterterms and is genuine signature of low energy chiral 
dynamics.  
\begin{equation}
I^{\rm (na)} = -{1\over 12\pi f^2}\cases{{2\over\pi} (m^2-\Delta^2)^{3/2} 
\sec^{-1}{m\over\Delta} + \Delta(\Delta^2 - {3\over2}m^2) {1\over\pi} 
\ln{m^2\over\mu^2}, &$\Delta\leq m$;\cr {-1\over\pi} (\Delta^2-m^2)^{3/2} 
\ln{\Delta-\sqrt{\Delta^2-m^2}\over \Delta+\sqrt{\Delta^2-m^2}} +
\Delta(\Delta^2 - {3\over2}m^2){1\over\pi}\ln{m^2\over\mu^2},&$\Delta\geq m$,} 
\end{equation}
where $\mu$ is the subtraction point.  
This nonanalytic contribution is plotted in Fig.~1, with $\mu$ set to be equal
to $\Delta$ and $2\Delta$.   
It is clear that, while $I^{\rm (na)}$ does behave like $m^3$ at large $m$ 
for both values of $\mu$, 
\begin{equation}
I^{\rm (na)} = -{1\over 12\pi f^2}\bigg[m^3 - {3\over2\pi}m^2\Delta
\ln{m^2\over\mu^2} + \dots\Bigg], \qquad m\gg\Delta, 
\end{equation}
there are nontrivial corrections in the chiral 
limit due to a non-vanishing $\Delta$ for $\mu=\Delta$, while the chiral 
behavior is much better-behaved with $\mu=2\Delta$.  
It should be understood that changes in $\mu$ generate extra analytic terms 
($m^2$ and $m^0$ terms in this case), which should be absorbed into 
$I^{\rm (a)}$.  
To see it more explicitly one can perform a Taylor expansion around $m=0$ 
\cite{JM}, 
\begin{equation}
I^{\rm (na)}=-{1\over12\pi f^2}{1\over\pi}\bigg[(-\Delta^3+{3\over2}\Delta 
m^2) \ln {\mu^2\over4\Delta^2} - {3\over 8}{m^4\over\Delta}\ln{m^2\over 
4\Delta^2} +{\cal O}({1\over\Delta^2})\bigg]
\end{equation}
The $\ln {\mu^2 / 4\Delta^2}$ term is actually analytic in $m^2$ which can be 
absorbed into the analytic part and canceled by counterterms.  
It vanishes, however, at $\mu=2\Delta$, explaining why the $\mu=2\Delta$ 
curve approaches the chiral limit so nicely.  
The leading nonanalytic contribution is $m^4 \ln m^2 \sim m_q^2 \ln m_q$, 
quite different from the $m^3$ behavior when $\Delta = 0$.  
In fact the $1/\Delta$ suppression factor is reminding us that this is a 
higher dimensional operator produced on integrating out the heavier 
intermediate state.  

The corresponding integral in Q$\chi$PT is 
\begin{equation}
I_q = {i\over f^2}\int {d^4p\over(2\pi)^4} {p^2\over v\cdot p - \Delta}
{-A_0 p^2 + M_0^2 \over (p^2 - m^2)^2},  
\end{equation}
where the subscript $q$ stands for ``quenched''.  
As discussed above, the quenched chiral singularities comes from the $M_0^2$ 
term.  
So for simplicity we will set $A_0 = 0$ is the discussion below, and 
\begin{equation}
I_q = {i M_0^2\over f^2}\int {d^4p\over(2\pi)^4} {p^2\over v\cdot p - \Delta}
{1 \over (p^2 - m^2)^2} = M_0^2 {dI\over dm^2} 
\end{equation}
We are only interested in the nonanalytic part $I_q^{\rm (na)}$, which is 
\begin{equation}
I_q^{\rm (na)} = -{M_0^2\over 12\pi f^2}\cases{-{3\Delta\over2\pi}
\ln{m^2\over\mu^2} + {3\over\pi} (m^2-\Delta^2)^{1/2} \sec^{-1}{m\over\Delta},
&$\Delta\leq m$;\cr -{3\Delta\over2\pi}\ln{m^2\over\mu^2} 
+ {3\over2\pi} (\Delta^2-m^2)^{1/2}\ln{\Delta-\sqrt{\Delta^2-m^2}\over 
\Delta+\sqrt{\Delta^2-m^2}},&$\Delta\geq m$,} 
\end{equation}
We see the expected $I\sim m\sim m_q^{1/2}$ nonlinear behavior for large $m$. 
\begin{equation}
I_q^{\rm (na)}=-{1\over12\pi f^2}\bigg[{3\over2}M_0^2m-{3\over2\pi}M_0^2\Delta 
\ln{m^2\over\mu^2} + \dots \bigg],\qquad m\gg\Delta.  
\end{equation}
And for small $m$, 
\begin{equation}
I_q^{\rm (na)}=-{M_0^2\over12\pi f^2}\Bigg[{3\Delta\over2\pi}
\ln{\mu^2\over4\Delta^2}-{3\over4\pi}{m^2\over\Delta}\ln{m^2\over4\Delta^2}
+{\cal O}({1\over\Delta^2})\Bigg], 
\end{equation}
so the leading nonanalytic term is $m^2\ln m^2 \sim m_q\ln m_q$, again quite 
different from the $m\sim m_q^{1/2}$ behavior with vanishing $\Delta$.  
Another interesting point is shown in Fig.~2, where $I_q^{\rm (na)}$ is 
plotted with $\mu=\Delta$ and $2\Delta$.  
In this case a change in $\mu$ just shifts the curve vertically but do not 
change the ``shape'' at all, as dictated by the Taylor expansion above, where 
the coefficient in front of the $\ln \mu^2$ term is just a constant in $m$.  
Note that such a vertical shift would not affect extrapolations to the 
chiral limit at all as the ``shapes'' of the curves are identical.  

What does the analysis above imply for real lattice QCD simulations?  
Note that the nonlinear behavior in Fig.~2 appears around $m/\Delta \sim 0.5$.
So the implication of Fig.~2 is twofold.  
Firstly, it suggests that, for $m>\Delta/2$, the extrapolation of hadron masses
to the chiral limit should be linear, a fact which is consistent with existing 
lattice results.  
On the other hand, it predicts the breakdown of such linear behaviors at  
$m\leq \Delta/2$, a regime which is not yet well-explored.  
In real lattice QCD simulation, the application of this analysis is 
complicated by the existence of more than one possible intermediate states, 
and hence more than one $\Delta$. 
However, one can still make some qualitative predictions.  
For example, since the N--$\Delta$ splitting is around 300 MeV, one expects 
the extrapolation for the nucleon mass departs from a linear behavior at 
around 150 MeV (about the physical pion mass).  
On the other hand, the small $\rho$--$\omega$ splitting (17 MeV) means that 
probably a simple linear extrapolation is sufficient in the vector meson 
sector.  
Certainly more detailed studies along this line, especially for physical 
quantities other than masses, will be very useful to the extraction of 
physical information from lattice data.  

\section{Hadron Decay in Quenched QCD}

In this previous section, we have studied two point functions with the 
intermediate state $X_i$ {\it heavier\/} than the external state $X_e$, 
{\it i.e.}, $\Delta=X_i-X_e>0$.  
In this section, we will consider the opposite, where $\Delta<0$ and the 
intermediate state is {\it lighter\/} than the external state.  
In particular, we are interested in the case which $\Delta < -m$, and at 
large Euclidean time the two point function will be dominated the $X_i\pi$ 
state, resulting in a cut in the two point function.  
However, often these decays involve creating a quark--antiquark pair from 
the vacuum.  
The question is, can hadron decay be studied on lattice under the quenched 
approximation?  

In this section, we are going to derive ``selection rules'' for the hadron 
decays under quenched approximation.  
For example, an obvious selection rule is {\it the daughter hadrons cannot 
contain any valence quark which does not appear in the mother hadron as 
valence quark}, as these new valence flavors have to from an internal quark 
loop.  
We will see below that quenching does impose strong constraints on the 
possible decay mechanisms.  
We will focus at system with the standard and hairpin axial current couplings 
(denoted by $g$ and $h$ respectively), but no other kind of couplings 
({\it i.e.}, $g_1$ and $g_3$ in Ref.~\cite{BCF} are out).  
Then there are two classes of diagrams representing hadron decays: 

(1)  Diagrams which all the vertices are standard couplings $g$, and no 
hairpin propagators.  
The corresponding integrals will be proportional to $g^2$.  
Since these decays have clear analogs in standard (unquenched) QCD, we will 
call them ``standard decays''.   

(2)  Decays with one hairpin axial vertex (integral $\sim gh$), and/or 
diagrams with the ``double pole'' propagator on the Goldstone boson line 
(integral $\sim g^2 M_0^2$).  
These decays may either have analogs in QCD or be merely quenched artifacts.  
We will call these ``hairpin decays''.  

We will first consider standard decays for meson.  
It is trivial to see that OZI suppressed decays, like $\phi\to\rho\pi$, 
cannot take place under the quenched approximation.  
However, not all OZI allowed decays can survive quenching.  
Decays through Fig.~3a, which contains an internal quark loop, are not 
allowed in quenched QCD, while decays through Fig.~3b are preserved.  
Note, however, that the quark and the antiquark from an external meson 
line in Fig.~3b annihilate, demanding that the external meson must be flavor 
neutral.  
However, isospin symmetry, which is a good symmetry of Q$\chi$PT, can rotate 
flavor neutral states into flavor states {\it unless the flavor neutral 
state has isospin zero}.  
Hence, we have the selection rule that for {\it the only possible standard 
meson decays are OZI allowed decays of $I=0$ mesons.}  
The amplitude of such decays in quenched QCD is identical to its counterpart 
in standard QCD.  
An example of such decay is $\omega\to\rho\pi$ in the vector meson sector, and
its tensor meson analog $f_2\to a_2\pi$;  
unfortunately these processes do not happen in the real world where the total 
mass of the daughter particles is larger than the mother particle.  
It turns out that light mesons, which form U(3) nonets, usually mix ideally so 
that the $I=0$ mass eigenstates are the $s\bar s$ state and the $u\bar u - 
d\bar d$ states.  
The $s\bar s$ states, like the $\phi$ vector meson and the $f'_2$ tensor 
meson, can either decay through OZI suppressed modes like $\phi\to\rho\pi$, 
which is clearly disallowed in quenched QCD, or modes like $\phi\to2K$, which 
is described by Fig.~3a and hence also do not survive quenching.  
And while the $u\bar u - d\bar d$ state, like the $\omega$ vector meson and 
the $f_2$ tensor meson in principle can decay through Fig.~3b, they are so 
light that they cannot decay into the $I=1$ state in the same nonet ($\rho$ 
and $a_2$, respectively).  
So the only possibility for standard decays in the meson sector are decays 
of an $I=0$ state from a highly excited nonet to two lighter $I=1$ states in 
other nonets.  
These decays cannot be described by Q$\chi$PT as they involve transitions 
between different nonets, and it is probably difficult to really observe these
decays on the lattice as these highly excited states may be very dirty.  
In passing we note that heavy mesons do not have standard decays in quenched 
QCD.  

In the baryon sector we are interested in studying decays like $\Delta\to
{\rm N}\pi$ and $\Sigma_c^{(*)}\to\Lambda_c\pi$.  
The former decay has been studied in detail in Ref.~\cite{LS}, from which we 
will borrow heavily.  
The main complication in studying these baryon decays is that the initial 
and final baryons have different isospin wave functions to satisfy the 
spin-statistics theorem.  
For concreteness, Ref.~\cite{LS} has chosen to study the decay of 
$\Delta^{++}$, which in standard QCD the only possible decay mode is: 
\begin{equation}
\Delta^{++} = uuu \to u\bar d + duu = \pi^+ + p^+
\end{equation}
Note that a $d \bar d$ pair is created.  
So {\it naively} $\Delta^{++}$ decay is prohibited by quenching.  
This argument, however, was shown to be wrong in Ref.~\cite{LS}.  

Standard decays for baryons can go through either Fig.~4a, which has an 
internal quark loop and hence does not survive quenching, and Fig.~4b, which 
does.  
We will denote the amplitude of $\Delta^{++}$ decaying through Fig.~4a as 
$\Gamma_q$, where $q$ is the quark on the internal loop, while $\Gamma_0$ is 
the amplitude of its decay through Fig.~4b.  
In the unquenched world, the total decay amplitude is
\begin{equation}
\Gamma^{\rm unquenched} = \Gamma_0 + \sum_q\Gamma_q = \Gamma_0 + \Gamma_u 
+ \Gamma_d.  
\end{equation}
On the other hand, fermionic statistics of the quark forbids the existence 
of an intermediate $uuu$ octet.  
Hence we have 
\begin{equation}
\Gamma_0 = - \Gamma_u ,
\end{equation}
and hence 
\begin{equation}
\Gamma^{\rm unquenched} = \Gamma_d = \Gamma_q.
\end{equation}

The $\Delta^{++}$ decay amplitude after quenching is 
\begin{equation}
\Gamma^{\rm quenched} = \Gamma_0 + \sum_{q,\tilde q}\Gamma_q = \Gamma_0 
+ \Gamma_u + \Gamma_{\tilde u} + \Gamma_d + \Gamma_{\tilde d}.  
\end{equation}
The cancelation of quark loops and ghost loops demands that 
\begin{equation}
\Gamma_q + \Gamma_{\tilde q} = 0, 
\end{equation}
which gives 
\begin{equation}
\Gamma^{\rm quenched} = \Gamma_0 = - \Gamma_q = - \Gamma^{\rm unquenched}.  
\end{equation}
So the decay amplitude changes sign under quenching.  
A more illuminating way to present the result is to note that 
\begin{equation}
\Gamma^{\rm quenched} = (\Gamma_0 + \Gamma_u) + \Gamma_{\tilde u} + (\Gamma_d 
+ \Gamma_{\tilde d}) = \Gamma_{\tilde u} = \Gamma_{\tilde q}, 
\end{equation}
as the sum in each pair of parentheses vanishes.  
So in quenched QCD $\Delta^{++}$ decays to $(u\skew2\bar{\tilde u})
(\tilde uuu)$,which clearly has the opposite sign as the QCD decay.  

How does this lesson generalize to other baryon decays?  
We will focus our attention to the orbitally unexcited baryons, which are 
the states with most physical interest.  
In other words, we will consider the octet-decuplet light baryon system, 
and the antitriplet-sextet heavy baryon system.  
Now we make the phenomenological observation that, while baryon decays 
through pion emission are possible in both of these systems ({\it e.g.}, 
$\Delta\to{\rm N}\pi$ and $\Sigma_c^{(*)}\to\Lambda_c\pi$), in neither of 
these system can baryon decay through $K$ or $\eta$ decay as physically 
these mesons are too massive.  
We will make the assumption that this pattern continue to hold in quenched 
QCD.  
Then Fig.~4b demands that, for decay through pion emission to take place, 
the initial state must has at least two light valence quarks.  
So the selection rule is: {\it standard decays for baryons are possible 
if the decaying baryon has at least two light valence quarks}, and the list 
includes $\Delta\to$ N and $\Sigma_{(c)}^{(*)}\to\Lambda_{(c)}$.  
And for all of these decays, the quenched amplitude has the same magnitude 
but opposite sign as the unquenched counterpart.  
On the other hand, orbitally unexcited $\Xi_{(Q)}$'s and $\Omega_{(Q)}$'s do 
not have standard decays under the quenched approximation.  
Our rule does not apply to orbitally excited baryons, which may have enough 
phase space to decay through $K$ and $\eta$ emission, but again studying 
these states on the lattice may be quite difficult.  

After studying the standard decays, let's now move on to the hairpin decays.  
Hairpin decays either have a ``double pole'' Goldstone propagator, leading 
to an integral proportional to $M_0^2 g^2$ (the hairpin axial coupling $h$ 
cannot appear; see Section 1), or are interference effects between the 
standard coupling $g$ and the hairpin coupling $h$, {\it i.e.}, proportional 
to $gh$.  
Note that the $h^2$ term is absent here as in that case the Goldstone boson 
line would form a quark loop.  
In other words, the hairpin vertex $h$ can cross interfere with $g$, but not 
self interfere, which is an evidence for non-unitarity.  

The important observation here is, hairpin decays always have the form 
$X_e\to X_i\eta'$, as only $\eta'$ couples through hairpin coupling, or 
propagate through the ``double pole'' propagator.  
Since $\eta'$ is a singlet under flavor SU(2) and SU(3), however, it means 
that {\it hairpin decays can take place only between multiplets with the 
same representation under the flavor group}.  

In the view of this result, the most natural places to look for these 
hairpin decays are transitions between a heavy quark doublet, like 
$D^*\to D\eta'$ and $\Sigma_c^*\to\Sigma_c\eta'$.  
Since the two states in a heavy quark doublet is related by just flipping 
the heavy quark spin, they always have identical flavor representation.  
As discussed above, there are no standard decay for $D^*$, and this hairpin 
decay is the only possible mode under the quenched approximation.  
The $\Sigma_c^*$ baryon can decay to $\Lambda_c\pi$, which would probably 
dominate over this hairpin mode, as the latter in suppressed in the large 
$N_c$ limit \cite{CR2}.  
In the real world, moreover, $\Sigma_c^*$ is only 80 MeV above $\Sigma_c$, 
so probably this hairpin decay cannot take place after all.  

The strange quark is not massive enough to be considered a heavy quark, and 
while the ``strange quark doublet'' $\Sigma^*$ and $\Sigma$ have the same 
representation under flavor SU(2) (both have $I=1$), their flavor SU(3) 
representations are different ($\Sigma$ is in the octet, $\Sigma^*$ is in 
the decuplet).  
Hence the hairpin decay $\Sigma^*\to\Sigma\eta'$ is flavor SU(3) suppressed.  
Lastly, note that hairpin decays from the tensor meson nonet to the vector 
meson nonet are allowed under this selection rule, and in principle one can 
study these decays on the lattice.  
These decays, however, are not described within the framework of (Q)$\chi$PT, 
where the tensor meson number and vector meson number are separately conserved.

In conclusion, we have chosen that quenching provides severe ``selection 
rules'' on hadron decays.  
Studying decay on lattice are important as it provide one of the most direct 
way to determine the values of the coupling constants in the Q$\chi$PT 
lagrangian (just as experimental decay rate can lead to determinations of 
coupling constants in the standard chiral lagrangian).  
One can obtained ${\cal C} \sim g_{\Delta{\rm N}\pi}$ \cite{LS} from 
$\Delta\to$N$\pi$, and $g_3 \sim g_{\Sigma_Q^{(*)}\Lambda_Q\pi}$ \cite{C} from 
$\Sigma_c^{(*)}\to\Lambda_c\pi$.  
Regretfully, hadron decays are not observable in most of the lattice QCD 
simulations in the present generation as the light quark masses are still 
relatively large, and hence $m > m_\Delta - m_{\rm N}$.  
But it is probable that, with the improvement in algorithms and increase in 
available computational power, probing the small $m$ region would be a 
feasible endeavor in the not-too-distant future.  

\acknowledgements

I would like to thank S.J.~Rey and S.~Sharpe for valuable discussions.  
This work is supported in part by the National Science Foundation.

\begin{figure}
\caption{Plot of $-12\pi f^2 I^{\rm (na)}/\Delta^3$ vs $m/\Delta$.  
The dashed curve is obtained with $\mu = \Delta$, while the solid curve is 
obtained with $\mu = 2 \Delta$.}  
\end{figure}

\begin{figure}
\caption{Plot of $-12\pi f^2 I^{\rm (na)}/(M_0^2\Delta)$ vs $m/\Delta$.  
The dashed curve is obtained with $\mu = \Delta$, while the solid curve is 
obtained with $\mu = 2 \Delta$.}  
\end{figure}

\begin{figure}
\caption{Diagrams contributing to two point functions of a decaying meson.  
Fig.~3a contains an internal quark loop and does not survive quenching, 
while Fig.~3b does not contain internal quark loop and contributes in 
quenched QCD.}
\end{figure}

\begin{figure}
\caption{Diagrams contributing to two point functions of a decaying baryon.  
Fig.~4a contains an internal quark loop and does not survive quenching, 
while Fig.~4b does not contain internal quark loop and contributes in 
quenched QCD.}
\end{figure}

\begin{references}
\bibitem{Morel} A.~Morel, J.~Physique {\bf 48} 111 (1987).  
\bibitem{S1} S.R.~Sharpe, Phys.~Rev.~{\bf D41} 3233 (1990).  
\bibitem{S2} S.R.~Sharpe, Phys.~Rev.~{\bf D46} 3146 (1992).  
\bibitem{BG1} C.W.~Bernard and M.F.~Golterman, Phys.~Rev.~{\bf D46} 853 (1992).
\bibitem{BG2} C.W.~Bernard and M.F.~Golterman, Phys.~Rev.~{\bf D49} 486 (1994).
\bibitem{BCF} M.~Booth, G.~Chiladze and A.F.~Falk, Phys.~Rev.~{\bf D55} 3092 
(1997).  
\bibitem{CR1} C.K.~Chow and S.J.~Rey, hep-ph/9708432 (1997).  
\bibitem{LS} J.N.~Labrenz and S.R.~Sharpe, Phys.~Rev.~{\bf D54} 4595 (1996). 
\bibitem{KK} M.~Kim and S.~Kim, hep-lat/9608091 (1996).  
\bibitem{CR2} C.K.~Chow and S.J.~Rey, in preparation. 
\bibitem{B} M.~Booth, Phys.~Rev.~{\bf D51} 2338 (1995).  
\bibitem{SZ} S.R.~Sharpe and Y.~Zhang, Phys.~Rev.~{\bf D53} 5125 (1996). 
\bibitem{SZ2} S.R.~Sharpe and Y.~Zhang, Nucl.~Phys.~{\bf B47} (Proc.~Suppl.) 
441 (1996).  
\bibitem{C} G.~Chiladze, hep-ph/9704426 (1997).  
\bibitem{PC} S.~Sharpe, private communication (1997).  
\bibitem{JM} E.~Jenkins and A.V.~Manohar, in {\sl Proceedings of the Workshop 
on ``Effective Field Theories of the Standard Model''} ed.~Ulf.~Mei\ss ner,
World Scientifc (1992).  
\end{references}
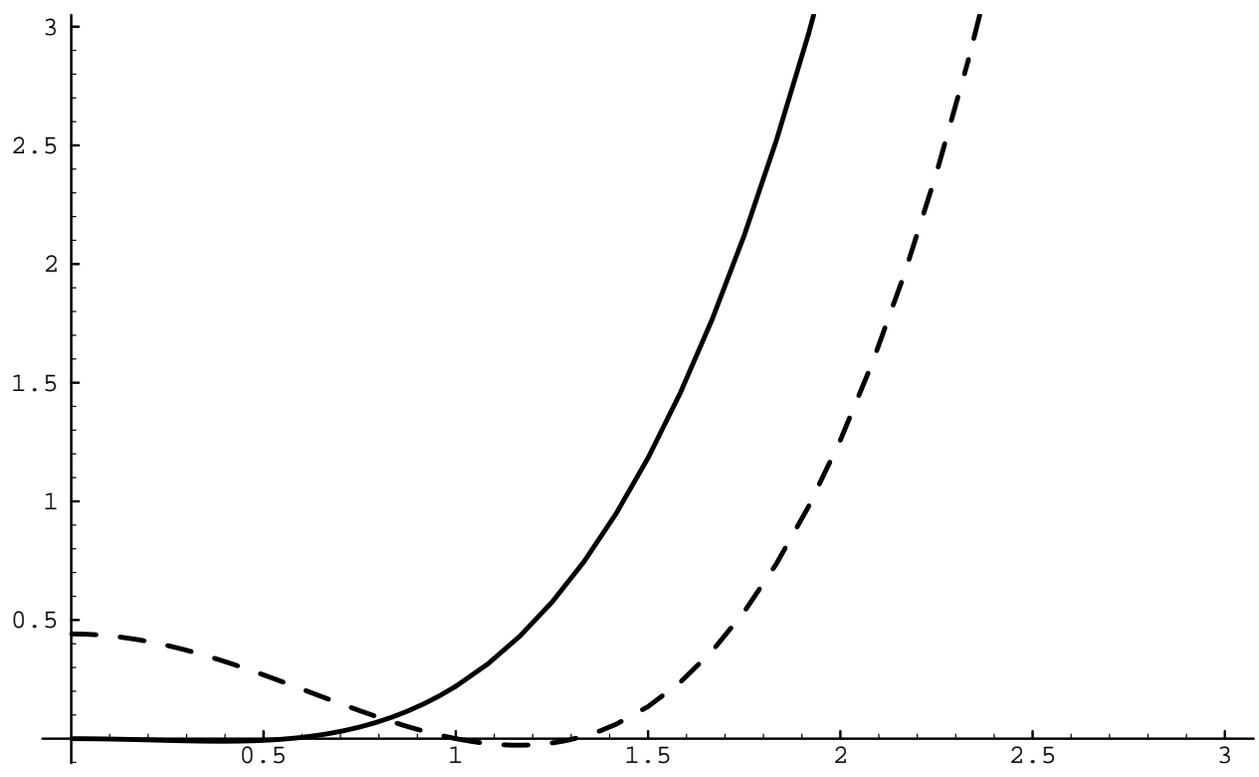
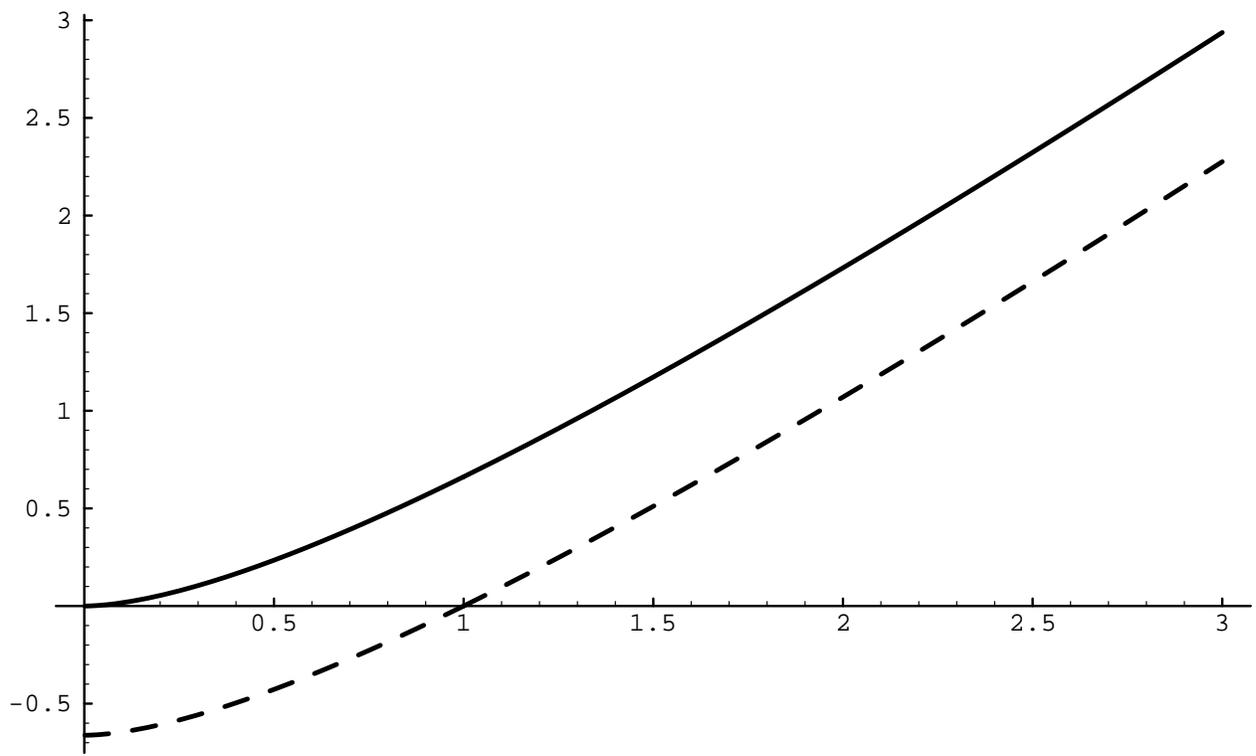
\end{document}